\pdfoutput=1
\documentclass[conference]{IEEEtran}
\usepackage{cite}
\usepackage{amsmath,amssymb,amsfonts}
\usepackage[noend]{algorithmic}
\usepackage{graphicx}
\usepackage{textcomp}
\usepackage{xcolor}
\def\BibTeX{{\rm B\kern-.05em{\sc i\kern-.025em b}\kern-.08em
		T\kern-.1667em\lower.7ex\hbox{E}\kern-.125emX}}

\usepackage{balance}  
\usepackage{ntheorem}
\usepackage{times}
\usepackage{epsfig,subfigure}
\usepackage[linesnumbered,ruled,noend]{algorithm2e}
\usepackage{graphicx,floatrow}
\usepackage{listings}
\usepackage{tikz}
\usepackage[T1]{fontenc}
\usepackage{pgfplots}
\usepackage[colorlinks,linkcolor=black,anchorcolor=black,citecolor=black]{hyperref}
\usepackage{booktabs}
\usepackage{multirow,multicol}
\usepackage{threeparttable}
\usepackage{comment}

\newcommand{\Paragraph}[1]{~\vspace*{-0.9\baselineskip}\\{\bf #1}}
\newcommand{\nop}[1]{}
\newtheorem{definition}{Definition}

\newcommand{\myproof}[1]{
	\begin{IEEEproof}
		#1
	\end{IEEEproof}
}

\newcommand{\optionshow}[2]{#1}

\newcommand{\picfolder}{./}

\renewcommand\baselinestretch{1}

\nop{
	
}

\usepackage{floatrow}
\newfloatcommand{capbtabbox}{table}[][\FBwidth]

\usepackage{soul}
\soulregister\cite7 
\soulregister\citep7 
\soulregister\citet7
\soulregister\ref7 
\soulregister\pageref7 

\begin{document}

	
	\title{Deep Analysis on Subgraph Isomorphism}
	
	\author{%
		{ {Li Zeng${^\dag}$}, Yan Jiang{${^\dag}$}, Weixin Lu{${^\dag}$}, Lei Zou{${^\dag}$}}%
		\\
		\fontsize{10}{10}\selectfont\itshape $~^{\dag}$Peking University, China;
		\\
		\fontsize{10}{10}\selectfont\ttfamily\upshape $~^{\dag}$$\{$li.zeng,yanjiang97,weixinlu,zoulei$\}$@pku.edu.cn
		\\}


	\maketitle
	
	\begin{abstract}
		Subgraph isomorphism is a well-known \emph{NP-hard} problem which is widely used in many applications, such as social network analysis and knowledge graph query. 
		Its performance is often limited by the inherent hardness.
		Several insightful works have been done since 2012, mainly optimizing pruning rules and matching orders to accelerate enumerating all isomorphic subgraphs.
		Nevertheless, their correctness and performance are not well studied.
		First, different languages are used in implementation with different compilation flags.
		Second, experiments are not done on the same platform and the same datasets.
		Third, some ideas of different works are even complementary.
        Last but not least, there exist errors when applying some algorithms.
		In this paper, we address these problems by re-implementing seven representative subgraph isomorphism algorithms as well as their improved versions,  and conducting comprehensive experiments on various graphs.
		The results show pros and cons of state-of-the-art solutions and explore new approaches to optimization.
	\end{abstract}
	

	
	
	
\section{Introduction}  \label{sec:introduction}

\begin{figure}[htbp]   
\centering
\includegraphics[width=8cm]{\picfolder 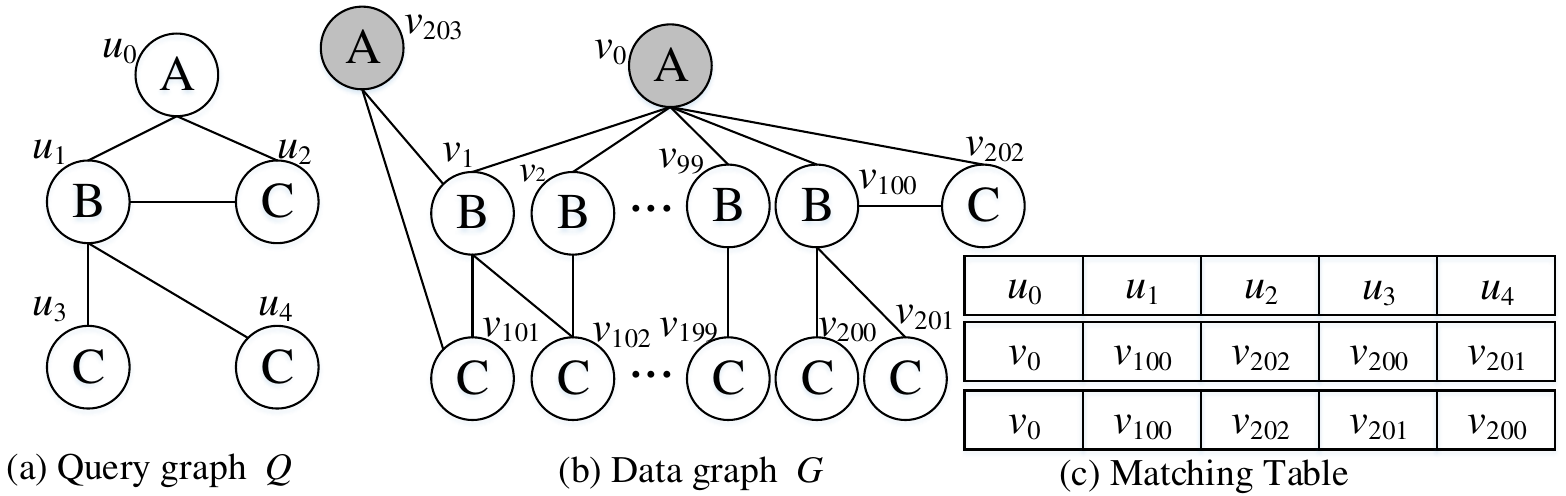}   
\vspace{-0.1in}          
\caption{An example of Query Graph and Data Graph}        
\label{fig:example}   
\end{figure}

Nowadays graphs are used to model many complicated structures and schema-less data, such as social networks, biological structures, and chemical compounds.
Many interesting problems arise from the broad application of graph data.
Among these, subgraph isomorphism is a fundamental problem: how to enumerate all subgraph isomorphism-based matches of a query graph $Q$ over a data graph $G$, which is the focus of this work.
A running example of $Q$ and $G$ is given in Figure \ref{fig:example}, where Figure \ref{fig:example}(c) illustrates the matching of $Q$ over $G$.
It is a well-known NP-hard problem \cite{npc} and has many important applications, such as chemical compound search, social network analysis and so on.

In big data era, subgraph isomorphism algorithms are continuously optimized to keep pace with growing graph size.
\cite{SubgraphIsomorphismComparison2012} conducted experimental analysis of works before 2012 to evaluate different pruning strategies and matching orders in different algorithms.
Experiments in \cite{SubgraphIsomorphismComparison2012} shows that QuickSI \cite{QuickSI} and GraphQL \cite{GraphQL} perform the best in general.

However, there are some factors not considered in \cite{SubgraphIsomorphismComparison2012}:social network datasets, various query types and memory consumption comparison.
Besides, great progresses are made after 2012, including TurboISO \cite{TurboISO}, BoostISO \cite{BoostISO}, CFL-Match \cite{CFL-Match}, VF3 \cite{VF3}, CECI \cite{CECI} and DAF \cite{DAF}.
These new solutions show higher ability and provide new insights on optimization.
Existing experiments fail to provide a comparison of them:
\begin{itemize}
\item Different languages are used with different compilation flags. For example, GraphQL is implemented in \textit{Java} and VF3 is compiled with \textit{-O6} option.
\item Experiments are not done on the same platform (Windows/Linux) and  the same datasets. Social networks are not used in several works and the influence of various query types is not well studied.
\item Ideas of different works are complementary. 
BoostISO proposes preprocessing techniques that can be integrated into all subgraph isomorphism algorithms.
\item Errors occur when applying VF3 and BoostISO.
The 2-hop pruning strategy of  VF3 only works for induced subgraph isomorphism.
Besides, the original BoostISO algorithm  yields wrong results in some cases.
\end{itemize}

To address problems listed above, we conduct new experiments that are rigorous and comprehensive.
To eliminate the interference of various factors (e.g., programming languages, implementer's programming skills, etc.), we try our best to re-implement these solutions using C++ on Linux.
They are compiled using the same compiler and options and run on the same platform.
Besides, during matching all contents are kept in memory, thus disk I/O is not considered.

Furthermore, we revisit BoostISO from two perspectives.
First, we evaluate the preprocessing techniques proposed in BoostISO by integrating them with existing subgraph isomorphism algorithms.
Experiments confirm the effectiveness of these techniques. 
Second, we find out the error of the original BoostISO algorithm that leads to wrong answers and performance deficiencies.
Thus, we customize BoostISO to ensure correctness and experimentally evaluate our modification.   

In summary, we made the following contributions.
\begin{itemize}
    \item We figure out the intrinsic innovations of state-of-the-art algorithms(QuickSI, GraphQL, TurboISO, BoostISO, CFL-Match, VF3, CECI and DAF) and propose a novel framework to present them. Besides, we fix some significant mistakes of VF3 and BoostISO.
    \item To compare these solutions fairly and empirically, we re-implement them in the same platform and compare them using the same benchmarks.
    \item Deep analysis is done on the experimental results to understand the pros and cons of different algorithms.
Through the analysis we report significant discoveries: 
    \begin{itemize}
        \item The ranking order of memory consumption is QuickSI, VF3, GraphQL, CFL-Match, TurboISO, CECI, DAF.
            This order corresponds to the size of indices in these algorithms.
        \item As for time performance, there is no single winner. But the overall top three solutions are VF3, QuickSI and DAF.
Tree-search algorithms (VF3 and QuickSI) leads the way in both time performance and memory consumption.
Newer index-based algorithms all adopt heavy filtering and complex pruning techniques, but the final performance is not convincing enough.
This means that a better tradeoff of pruning ability and pruning overhead  should be explored in the future.
        \item The effect of BoostISO is rather limited, but it can help improve some bad cases.
New progress should be made to decide whether to use BoostISO or not in a specific situation.
    \end{itemize}
\end{itemize}

The remainder of this paper is organized as follows. 
Section \ref{sec:background} reviews the background.
Details of our implementations are given in Section \ref{sec:implementation}.
The techniques of BoostISO is thoroughly studied in Section \ref{sec:optimization}.
Section \ref{sec:experiment} presents experimental results and Section \ref{sec:conclusion} concludes the paper.

\section{Background}  \label{sec:background}

\subsection{Problem Definition} \label{sec:problem}

\begin{definition}\label{def:graph} \textbf{(Graph)}
    A graph is denoted as $G=\{V,E,L\}$, where $V$ is the set of vertices; $E \subseteq V \times V$ is the set of undirected edges; $L$ is a labeling function that maps a vertex (of $V(G)$) to a label.
The label function of $G$ can also be specified as $L_{G}$.
$V(G)$ and $E(G)$ are used to denote vertices and edges of graph $G$, respectively.
\end{definition}

\begin{definition}\label{def:subgraph} \textbf{(Subgraph)}
	Given a graph $G=\{V,E,L\}$, a subgraph of $G$ is denoted as $G^{\prime}=\{V^{\prime},E^{\prime},L^{\prime}\}$, where vertex sets $V^{\prime}$ and edge sets $E^{\prime}$ in $G^{\prime}$ are subsets of $V$ and $E$, respectively, denoted as $V^{\prime} \subseteq V$ and $E^{\prime} \subseteq E$. Furthermore, for vertex label functions, $L^{\prime} \subseteq L$.
\end{definition}

\begin{definition}\label{def:inducedSubgraph} \textbf{(Induced Subgraph)}
    Given a graph $G=\{V,E,L\}$, $G^{\prime}$ is called an induced subgraph of $G$ iff:
    (1) $G^{\prime}$ is a subgraph of $G$ according to Definition~\ref{def:subgraph};
    (2) $\forall v_{1},v_{2}\in V^{\prime}$, $\overline{v_{1}v_{2}}\in E \longrightarrow \overline{v_{1}v_{2}}\in E^{\prime}$.
\end{definition}

\begin{definition}\label{def:graphisomorphism} \textbf{(Graph Isomorphism)}
	Given two graphs $H$ and $G$, $H$ is \emph{isomorphic} to $G$ if and only if there exists a bijective function $f$ between the vertex sets of $G$ and $H$ (denoted as $f:V(H)\longrightarrow V(G) )$, such that 
	\begin{itemize}
		\item $\forall u \in V(H), f(u) \in V(G) $ and $L_{H}(u)=L_{G}(f(u))$.
		\item $\forall  \overline{u_1u_2} \in E(H)$,  $\overline{f(u_1)f(u_2)} \in E(G)$.
        \item $\forall  \overline{u_1u_2} \in E(G)$,  $\overline{f^{-1}(u_1)f^{-1}(u_2)} \in E(H)$.	
	\end{itemize}
\end{definition}

\begin{definition}\label{def:subgraphisomorphism}\textbf{(Subgraph Isomorphism Search)}
	Given query graph $Q$ and data graph $G$, the subgraph isomorphism search problem is to find out all \emph{subgraph}s $G^{\prime}$ of $G$ such that  $G^{\prime}$ is \emph{isomorphic} to $Q$. $G^{\prime}$ is called a \emph{match} of $Q$.
\end{definition}

\begin{definition}\label{def:subgraphhomomorphism} \textbf{(Subgraph Homomorphism)}
    Given query graph $Q$ and data graph $G$, the subgraph homomorphism problem  is to find out all \emph{subgraph}s $G^{\prime}$ of $G$ such that a surjective function $f$ exists between the vertex sets of $Q$ and $G^{\prime}$ (denoted as $f:V(Q)\longrightarrow V(G^{\prime}) )$ satisfying:
	\begin{itemize}
        \item $\forall u \in V(Q), f(u) \in V(G^{\prime}) $ and $L_{Q}(u)=L_{G^{\prime}}(f(u))$.
        \item $\forall  \overline{u_1u_2} \in E(Q)$,  $\overline{f(u_1)f(u_2)} \in E(G^{\prime})$.
	\end{itemize}
\end{definition}

\nop{\begin{definition}\label{def:sm} \textbf{(Subgraph Isomorphism)}
Given $Q=\{V_{1},E_{1},L_{1}\}$ and $G=\{V_{2},E_{2},L_{2}\}$, a subgraph isomorphism (also called an \emph{embedding} or a \emph{match}) between $Q$ and $G$ is an injective function $M:V_{1}\rightarrow V_{2}$ such that (1) $\forall u \in V_{1}, L_{1}(u)=L_{2}(M(u))$, and (2) $\forall (u_{i},u_{j})\in E_{1}, (M(u_{i}),M(u_{j}))\in E_{2}$.
\end{definition}
\begin{definition}\label{def:problem} \textbf{(Problem Statement)}
Given a query graph $Q$ and a data graph $G$, the subgraph isomorphism search problem is to enumerate all distinct embeddings of $Q$ in $G$.
\end{definition}
}

Definition~\ref{def:graphisomorphism} is also called monomorphism \cite{GraphMatchSurvey}, while  Definition~\ref{def:subgraphisomorphism} is called \emph{embedding subgraph isomorphism}.
There is another type called \emph{induced subgraph isomorphism}, which requires that each subgraph $G^{\prime}$ in Definition~\ref{def:subgraphisomorphism} to be an induced subgraph (see Definition~\ref{def:inducedSubgraph}) of $G$.
Besides, \emph{subgraph homomorphism} (see Definition~\ref{def:subgraphhomomorphism}) allows that two query vertices are mapped to the same data vertices, which is forbidden in subgraph isomorphism.

An example of $Q$ and $G$ is given in Figure~\ref{fig:example}.
Only one match exists in this case, i.e., $M_{1}=\{(u_{0},v_{0}),(u_{1},v_{100}),(u_{2},v_{201}),(u_{3},v_{200})\}$.
Definition~\ref{def:subgraphisomorphism} is naturally supported by all subgraph isomorphism algorithms.
Though these solutions can be easily extended to process directed graphs, edge labels or label sets, that is not our focus.

This paper aims to compare the state-of-the-art algorithms of subgraph isomorphism search (Definition~\ref{def:subgraphisomorphism}).
Without loss of generality, we assume $Q$ is connected; otherwise, we can regard each connected component of $Q$ as a separate query and execute them individually.
Unless otherwise specified, we use $v$, $u$, $N(v)$, $C(u)$, $deg(u)$, $num(L)$, and $|A|$  to denote a data vertex, a query vertex, all neighbors of $v$, candidate vertices of $u$, degree of $u$, the number of currently valid elements in set $L$, and the size of set $A$, respectively.

\subsection{Related Work} \label{sec:related}

Most subgraph isomorphism algorithms follow some form of tree search with backtracking paradigm \cite{GraphMatchSurvey}, a depth-first search method that finds solutions in a memory efficient manner. 
Generally, during each step a partial match (initially empty) is iteratively expanded by adding to it new pairs of matched nodes; the pair is chosen using some necessary conditions that ensure the subgraph isomorphism constraints with respect to the nodes mapped so far, and usually also using some heuristics to prune as early as possible unfruitful search paths. 
Eventually, either the algorithm finds a complete match, or it reaches a state where the current partial mapping cannot be further expanded because of the matching constraints. 
Then, the algorithm will backtrack to probe other search paths.
Algorithm~\ref{alg:generic0} summarizes this process.


\Paragraph{Pure Tree Search Algorithms}.
Ullmann \cite{DBLP:journals/jacm/Ullmann76}, VF2 \cite{VF2} and VF3 \cite{VF3}  strictly follow the routine above. 
The main difference lies in the heuristic rules used for pruning states that donot lead to complete matches.
A common tradeoff is between the effectiveness of the pruning rules and the computational cost required for the evaluation of the pruning rules.
Backtracking paradigm has the advantage of not requiring all the states to be kept in memory: the maximum number of allocated states is proportional to the number of nodes, and thus if the space for each state is constant, this paradigm has a linear space complexity with respect to the number of data vertices.
However, their performance is still limited by the exponential search space because they fail to fully utilize the neighborhood restrictions of query structure. 
Besides, VF2 and VF3 adopt the definition of induced subgraph isomorphism, which enables several powerful pruning strategies.
In our context, optimization techniques targeting at induced subgraph isomorphism  can not be used.

\begin{algorithm}
    \label{alg:generic0}
    \caption{Tree-search framework}
    \KwIn{query graph $Q$, data graph $G$}
    \KwOut{all matches of $Q$ in $G$}
    \SetKwFunction{proc}{SubgraphSearch}
    \LinesNumbered
\nl      $M:=\emptyset$   \\
\nl      \ForEach{vertex $u$ in $V(Q)$}
    {
        \nl          collects candidates $C(u)$ \label{algcmd:filter0} \\
\nl           \textbf{if} $C(u)=\emptyset$ \textbf{then}  \KwRet \\
    }
\nl      call \proc($Q$,$G$,$M$,...)  \label{algcmd:backtrack}   \\ 
    \setcounter{AlgoLine}{0}
    \SetKwProg{myproc}{Procedure}{}{}
    \myproc{\proc{$Q$,$G$,$M$,...}}{
   \nl                \textbf{if} $|M|=|V(Q)|$ \textbf{then}  output successful match $M$ \\
\nl          \Else{
    \nl          	$u:=$NextQueryVertex(...) \label{algcmd:select0}  \\
      \nl           $C_{R}:=$RefineCandidates($M$,$u$,$C(u)$,...)  \\
         \nl      \ForEach{$v\in C_{R}$ and  $v$ is not matched}{
             \nl         \If{IsJoinable($Q$,$G$,$M$,$u$,$v$,...)}{ \label{algcmd:join0}
             \nl       // $\forall (u^{\prime},v^{\prime})\in M (\overline{uu^{\prime}}\in E(Q)\longrightarrow \overline{vv^{\prime}}\in E(G))$ \\
          \nl           UpdateState($M$,$u$,$v$,...) \\
          \nl           \proc($Q$,$G$,$M$,...)  \\
          \nl           RestoreState($M$,$u$,$v$,...)   \\
                }
            }
        }
    }
\end{algorithm}

\Paragraph{Index-based Algorithms}.
Except for tree search methods, current solutions \cite{QuickSI, GADDI, SPath, GraphQL, TurboISO, BoostISO, CFL-Match, CECI, DAF} mainly adopt filtering-and-verification framework and leverage some auxiliary indices to accelerate the exploration.
Two critical points are candidates pruning strategy (in filtering phase, see Line~\ref{algcmd:filter0}) and matching order selection (in verification, see Line~\ref{algcmd:select0}).
To select a good matching order, QuickSI \cite{QuickSI} gives priority to infrequent query vertices/edges, with the frequency defined by the number of occurrence in data graph $G$.
Note that QuickSI is originally designed for graph database, and it does not have filtering phase and auxiliary indices for matching on a single data graph.
However, choosing a more selective query edge can lead to a significantly less selective query path, thus SPath \cite{SPath} estimates the selectivity of each path of query $Q$, then it processes infrequent paths first.
Besides, SPath exploits neighborhood-based signature for filtering, in order to reduce $|C(u)|$.
To further minimize candidate sets, GraphQL \cite{GraphQL} performs pseudo-isomorphism test, which generates BFS trees for $u$ and $v\in C(u)$, checks if the containment relationship exists between the two, and iterates this process by enlarging the BFS trees.
Therefore, GraphQL mainly adopts two techniques for filtering: neighborhood-based signature and pseudo-isomorphism test.
These two pruning strategies are costly but efficient.
As for the matching order, GraphQL uses a simple greedy strategy.
Each time it selects an unmatched query vertex $u$ that is connected to already matched query vertices and that has the minimum candidate set.
In addition, SPath and GraphQL both have the problem of tuning parameters.
Their performance is very sensitive to parameters, but it is hard to decide optimal values for these parameters for each query.
The state-of-the-art algorithms (TurboISO, BoostISO, CFL-Match, CECI and DAF) also use the same framework and build auxiliary indices, which will be detailed in Section~\ref{sec:implementation}.

\nop{
\Paragraph{BFS-based Algorithms}.
There exists some works which use relational table join instead of depth-first graph exploration, thus we call them BFS-based algorithms.
They may consume much more memory to store intermediate tables, but they can naturally fit on parallel and distributed environments to process billion-node graphs.
Vertex-based solutions \cite{gStore, DBLP:journals/fcsc/ZengZ18} filter candidates for vertices first and join them according to the query structure.
In contrast, edge-based solutions \cite{RDF-3X, BitMat, peng2016answering} filter candidates for query edges first and then a natural table join is performed on these tables to generate matches.
There is another type, \emph{STwig}-based solution \cite{Trinity}, which is implemented in the memory cloud of Trinity \cite{Trinity}.
It first decomposes a query into a set of basic query units called STwigs and obtains all partial matches for each STwig. 
Later, a block-based pipeline join is performed to merge these intermediate results, with the join order given by a sampling-based strategy.
}
\nop{
Then, it obtains all possible intermediate results for each STwig. 
Here, a large amount of them must be transferred to other servers before join. 
Next, it performs block-based pipeline joins to merge all intermediate results. 
To define a join order, it uses a sampling based join optimization. 
However, it is unclear whether this algorithm is superior to the existing ones, since it
has not been empirically compared with them.
}

\Paragraph{Variants}.
These include homomorphism \cite{RDF-3X, TurboHom++, Trinity, gStore, DBLP:journals/fcsc/ZengZ18}, induced subgraph isomorphism (used in VF2 and VF3), and graph database search (checking the existence of a  query graph in a list of data graphs) \cite{QuickSI, CFQL}.
\cite{GraphMatchSurvey} presents details of these variants.
Approximate subgraph matching problems \cite{Ness,SIGMA,TALE} define matches with their own similarity measures.
Graph simulation \cite{fan2014querying, ma2014strong} can also be classified into this category.

\section{Implementation}\label{sec:implementation}



In this section, we introduce five state-of-the-art algorithms and give their implementation details.
Most of them follow the framework in Algorithm~\ref{alg:generic}, which consists of filtering phase and joining phase (also called verification).
Firstly, a tree structure $T$  is extracted from the query graph $Q$ (Line~\ref{algcmd:tree}).
The edges of $T$ are called tree edges $TE$, while the edges of $E(Q)\setminus E(T)$ are called non-tree edges $NTE$.
Secondly, according to the tree structure, candidates are generated for query vertices (Line~\ref{algcmd:filter}).
Generally, these candidates together with their linking information, are stored in some auxiliary index $I$.
Thirdly, the index $I$ is refined by the restrictions of tree edges or non-tree edges (Line~\ref{algcmd:refine}).
Then a matching order is determined by some cost model (Line~\ref{algcmd:order}).
Finally, an exploration is performed on the index $I$ to find all valid matches (Line~\ref{algcmd:join}).
The final step is done by backtracking, similar to the \emph{SubgraphSearch} procedure in Algorithm~\ref{alg:generic0}.
The former three steps are called ``filtering'', while the latter two are called ``verification''.

In Algorithm~\ref{alg:generic}, the spanning tree is generated for filtering  because tree is the simplest structure of a connected graph.
If there does not exist non-tree edges in $Q$, after filtering (Line~\ref{algcmd:filter}) $I$ contains exactly all correct answers.
By using a tree instead of the original query graph, a good balance is made between the cost of filtering and the filter efficiency.
Generally, the spanning tree is generated by performing breadth-first search from a selected source.
This is a simple but effective strategy.
Besides, the matching order is also based on the tree.

In subsections below, we will present the ideas of each work and their differences based on Algorithm~\ref{alg:generic0} and~\ref{alg:generic}.
QuickSI and GraphQL are older algorithms and  have been introduced in Section~\ref{sec:related}, thus we do not detail them here.
BoostISO is further studied in Section~\ref{sec:optimization} because it is  a general optimization independent from all solutions.

\begin{algorithm}
	\caption{New Generic framework}
	\label{alg:generic}
	\KwIn{query graph $Q$, data graph $G$}
	\KwOut{all matches of $Q$ in $G$}
\nl 	Rewrite $Q$ into a query tree $T$  \label{algcmd:tree} \\
\nl 	Find candidates for query vertices and build auxiliary indices ($I$) according to tree edges $TE$ of $T$ \label{algcmd:filter}    \\ 
\nl 	Refine candidates and update $I$ according to tree edges $TE$ or non-tree edges $NTE$ \label{algcmd:refine}  \\
\nl 	Determine matching order of query vertices \label{algcmd:order}  \\
\nl 	search in $I$ by backtracing \label{algcmd:join}   \\ 
\end{algorithm}

\subsection{TurboISO Algorithm} \label{sec:turboiso}


Based on Algorithm~\ref{alg:generic}, TurboISO adds two optimizations: neighborhood equivalence class (NEC) and candidate region (CR) division.
The first intuition is merging similar query vertices to eliminate duplicate computation caused by automorphism.
Automorphism means that a graph can be isomorphic to itself and there exists several matches.
Query vertices sharing common neighborhood structures can be combined into one NEC vertex $u^{\prime}$, then we find valid matches  for $u^{\prime}$.
When all NEC vertices are matched, for each NEC vertex $u^{\prime}$, we enumerate permutations of the real query vertices of $u^{\prime}$.
The second intuition is dividing candidate regions based on the data vertices matched to the first query vertex $u_{s}$ (denoted $u_{s}^{\prime}$ in $Q^{\prime}$).
Candidate regions can be searched one by one, and different regions can be explored in different matching orders.

\begin{figure}[htbp]   
	\centering
	\includegraphics[width=8cm]{\picfolder 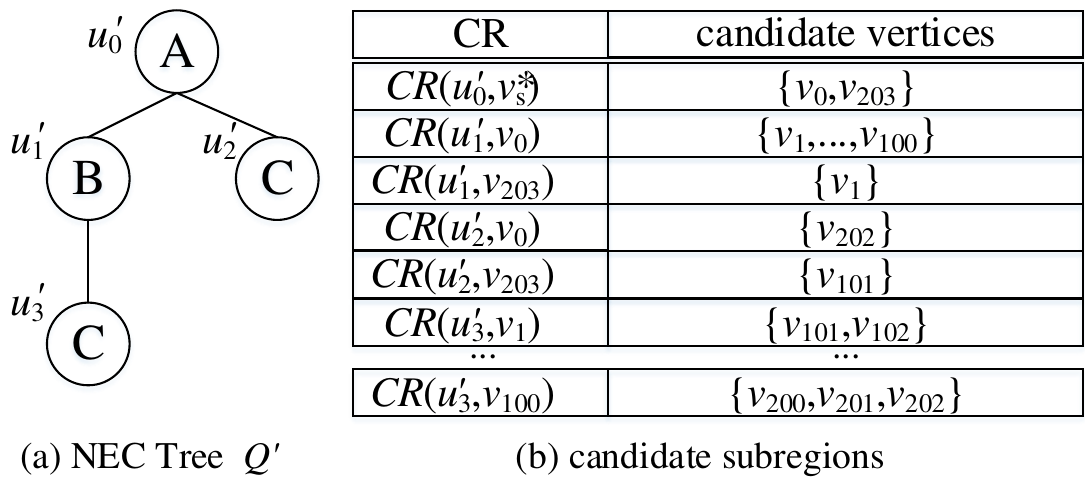}   
	\vspace{-0.1in}          
	\caption{Example of TurboISO}        
	\label{fig:turboiso}   
\end{figure}

Figure~\ref{fig:turboiso} shows the structure of TurboISO corresponding to Figure~\ref{fig:example}.
NEC vertices are found during the generation of query tree $Q^{\prime}$ (Line~\ref{algcmd:tree}).
$Q^{\prime}$ is also called NEC tree in this situation, and Figure~\ref{fig:turboiso}(a) shows an example ($u_{3}$ and $u_{4}$ are combined into one NEC vertex $u_{3}^{\prime}$).
Next, all candidate regions are explored according to the tree structure of $Q^{\prime}$ and candidates are stored in CR tree structure (Line~\ref{algcmd:filter}.)
Figure~\ref{fig:turboiso}(b) shows these regions (started from $v_{0}$ and $v_{203}$, respectively), where $CR(u^{\prime},v)$ denotes data vertices that are children of $v$ and matched to $u^{\prime}$.
Refine process (Line~\ref{algcmd:refine}) is not included in TurboISO\@.
To determine the matching order of each candidate region (Line~\ref{algcmd:order}), TurboISO performs candidate region exploration to count the number of candidate vertices for a given path of $Q^{\prime}$.
It favors the path with the minimum candidate set among the paths not selected yet, and each iteration an entire path is joined via backtracking.
Finally, for each region, a depth-first search is performed on CR tree to join candidates according to the restrictions of non-tree edges in $Q^{\prime}$ (Line~\ref{algcmd:join}).
During backtracking, for each NEC vertex $u^{\prime}$, a combination of data vertices (denoted $MC(u^{\prime})$) is formed for all real query vertices corresponding to $u^{\prime}$.
Only when a complete match is found, we enumerate all permutations of vertices in $MC(u^{\prime})$ to generate matches for the original query $Q$.
For example, a complete match of  $Q^{\prime}$ is $\{(u_{0}^{\prime},\{v_{0}\}),(u_{1}^{\prime},\{v_{100}\}),(u_{2}^{\prime},\{v_{202}\}),(u_{3}^{\prime},\{v_{200},v_{201}\})\}$, and permutations of $\{v_{200},v_{201}\}$ are enumerated for $u_{3}$ and $u_{4}$ to form two answers in Figure~\ref{fig:example}(c).

\Paragraph{Analysis}.
The space cost of CR tree is very high, i.e., $O(|E(G)|^{|V(Q)|-1})$.
The reason is that  many duplicate regions exist, e.g., $v_{0}$ and $v_{203}$ can be both extended to $v_{1}$, thus there are two identical $CR(u_{3}^{\prime},v_{1})$ regions.
This also brings much redundant computational work.
The process of finding NECs has a $|V(Q)|\times |E(Q)|$ time complexity, which is very lightweight.
As for Line~\ref{algcmd:order}, its time complexity is $O(|E(G)|^{|V(Q)|-1})$, because all CR states must be counted.


\subsection{CFL-Match Algorithm} \label{sec:cflmatch}

Based on Algorithm~\ref{alg:generic}, CFL-Match proposes two techniques to postpone cartesian products: the core-forest-leaf decomposition and the compact path-based index (CPI). 
Before the construction of BFS tree, the query graph is decomposed into three substructures: core, forest and leaf.

\begin{definition}\label{def:Core} \textbf{(Core Structure)}
    The minimal connected subgraph $Q_{c}$ of $Q$ that every vertex of $Q_{c}$ has at least two neighbors in $Q_{c}$. 
\end{definition}

\begin{definition}\label{def:Forest} \textbf{(Forest Structure)}
    The subgraph $Q_{f}$ of $Q$ consisting of all other edges not in $Q_c$.
\end{definition}

\begin{definition}\label{def:Leaf} \textbf{(Leaf Structure)}
	The subgraph $Q_l$ of $Q$ consisting of all degree-one vertices.
\end{definition}

After the decomposition of query graph, the matching is conducted by $Q_c\rightarrow Q_f\rightarrow Q_l$ order.
Three substructures are with decreasing graph density. 
In this way, cartesian products are postponed and number of unpromising partial mappings can be significantly reduced. 

\begin{figure}[htbp]   
	\centering
	\includegraphics[width=8cm]{\picfolder 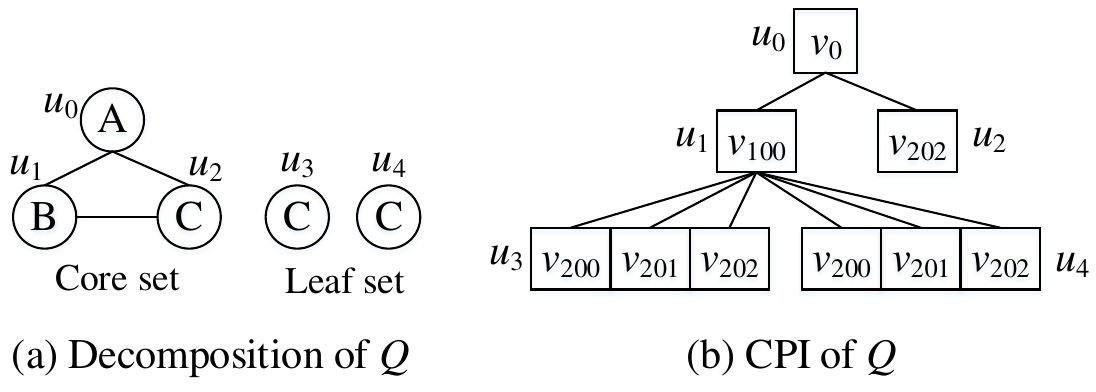}   
	\vspace{-0.1in}          
	\caption{Example of CFL-Match}        
	\label{fig:cfl-CPI}   
\end{figure}

Figure~\ref{fig:cfl-CPI}(a) shows the result of decomposition of query graph $Q$ in Figure~\ref{fig:example}. 
The core set is ${u_0, u_1, u_2}$ and the leaf set is ${u_3, u_4}$. 
Because all vertices are already covered by core and leaf set, the forest set is left empty. 

Vertex candidate and tree edge candidate sets are stored in CPI structure.
The  two sets are both constructed by neighbor set intersections and neighbor degree filtering.
First, a top-down construction is performed to generate CPI structure, corresponding to Line~\ref{algcmd:filter} of Algorithm~\ref{alg:generic}.
Second, a bottom-up refinement further prunes unpromising candidates.
During filtering, CFL-Match exploits both tree edges and non-tree edges as well as both directions of these edges to refine
candidates for each query vertex.


For example, Figure~\ref{fig:cfl-CPI}(b) shows the CPI constructed over $Q$ and $G$ in Figure~\ref{fig:example}. 
A vertex candidate set $C(u_i)$ is maintained for each query vertex $u_i$.
The line between $v_{100}$ in $C(u_1)$ and $v_{200}$ in $C(u_3)$ indicates that query edge $(u_1, u_3)$ can be matched to 
data edge $(v_{100}, v_{200})$.

As for matching order selection, each time CFL-Match selects the path in the BFS tree of $Q$ with the minimum cost function value, and adds all vertices of that path into matching order, until all vertices in $Q$ has been added. 
Here the cost function of path $\pi$ is defined by the estimated number of matches of $\pi$ together with the chosen paths $P$. 
Dynamic programming and greedy strategy are used in the estimation process, whose detail can be found in \cite{CFL-Match}. 
By choosing the path that generates minimum partial matches, large number of unpromising intermediate results are avoided. 
Note that paths selections are conducted within the core or forest set respectively, thus a vertex in core always precedes a vertex in forest in the matching order.
For example, in Figure~\ref{fig:cfl-CPI}(b), there are two paths in core set of $Q$, where $\pi_1=\{u_0, u_1\}, \pi_2=\{u_0, u_2\}$. $\pi_2$ has a smaller estimated number of matches, therefore the match order in core set is $\{u_0, u_2, u_1\}$.

\nop{
The matching process of leaf set are quite like techniques in TurboISO\@.
The leaf vertices with the same parent and label are merged into a single NEC\@.
After sorted in increasing order according to their number of candidates, each NEC $u^{\prime}$ is mapped to a combination of vertices from its respective $C(u^{\prime})$. 
Finally, we permute the mappings for query vertices in each NEC to obtain all matches.
Continuing the above example, we have $u_3$ and $u_4$ in the same NEC, which leads to two matches $\{(u_3, v_{200}), (u_4, v_{201})\}$, $\{(u_3, v_{200}), (u_4, v_{201})\}$
}

\Paragraph{Analysis}.
The space cost of CPI is $O(|V(Q)|\times |E(G)|)$.
The decomposition and CPI construction process take time $O(|E(Q)|)$ and $O(|E(Q)|\times |E(G)|)$ respectively, and matching order selection takes time $O(|E(Q)|\times |E(G)|)$ at most.


\subsection{VF3 Algorithm} \label{sec:vf3}

VF3 is a tree search method, which strictly follows Algorithm~\ref{alg:generic0}.
It is based on VF2 and leverage several novel techniques.
First, each query vertex $u$ is ranked with a probability estimator $P(u)=Pr(L(v)=L(u),deg(v)\geq deg(u))$.
Next, the matching order is decided by considering the node mapping degree (the number of edges between $u$ and already mapped nodes), the estimated probability, and the node degree.
Based on this order, state structures (the already mapped query vertices that $u$ is connected to) of $Q$ are precalculated to avoid duplicate computation in backtracking.
Then all vertices in $Q$ and $G$ are classified into different categories.
For example, vertices can be classified by the label and each vertex label represents one category.
Finally, a depth-first search is performed on data graph $G$, leveraging the same pruning rules as VF2 for each category.

\begin{figure}[htbp]   
	\centering
	\includegraphics[width=8cm]{\picfolder 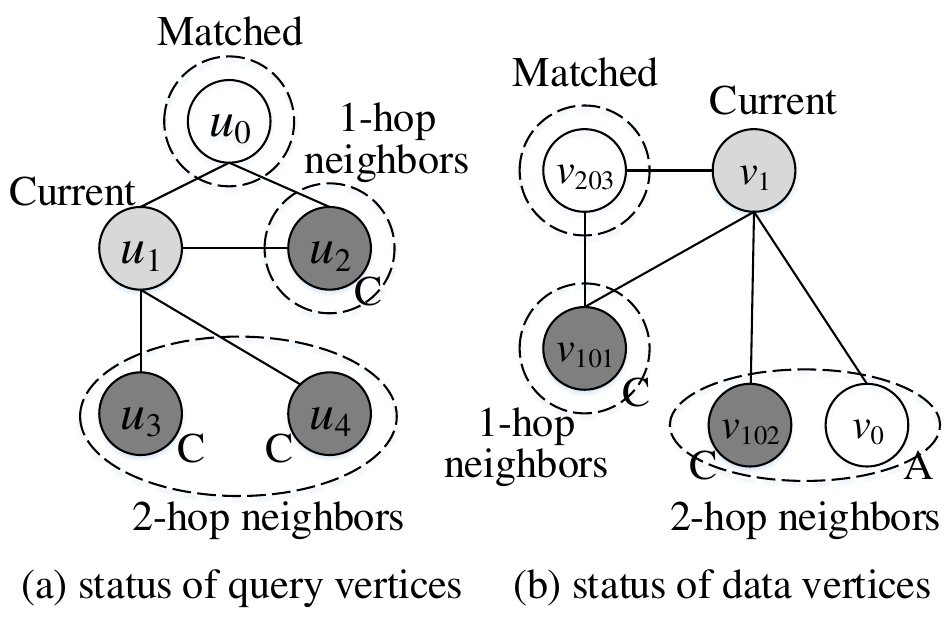}   
	\vspace{-0.1in}          
	\caption{Feasibility check of adding $(u_{1},v_{1})$}        
	\label{fig:vf3}   
\end{figure}

\Paragraph{Feasibility Check}.
To check if a new couple $(u_{c},v_{c})$ is feasible, VF3 verifies 1-hop and 2-hop consistence.
Figure~\ref{fig:vf3} shows an example of feasibility check, where $v_{203}$ is already mapped to $u_{0}$, and we try to map $v_{1}$ to $u_{1}$.
Nodes with deep color belong to category ``C'', nodes with light color belong to category ``B'' and nodes without color belong to category ``A''.
Let $l_{1}$ and $l_{1}^{\prime}$ be the set of vertices (1-hop neighbors) connected to already mapped nodes in $Q$ and $G$ respectively,  VF3 requires that $|l_{1}^{\prime}|\geq |l_{1}|$.
In Figure~\ref{fig:vf3}, the 1-hop neighbors in $Q$ and $G$ is $u_{2}$ and  $v_{101}$ respectively, thus the requirement is satisfied.
Another requirement of VF3 is $|l_{2}^{\prime}|\geq |l_{2}|$, where $l_{2}$ is the set of vertices (2-hop neighbors) connected to $u_{c}$ but not connected to already mapped vertices (the definition of $l_{2}^{\prime}$ is similar).
For category ``C'' in Figure~\ref{fig:vf3}, this check fails and we know that $(u_{1},v_{1})$ is invalid.

\Paragraph{Error of 2-hop pruning}.
However, the judgement of 2-hop pruning is not always true.
Assume that $v_{0}$ also belongs to category ``C'' and there is an edge between $v_{0}$ and $v_{203}$, then the modified example contains a complete match, but it still fails the check of 2-hop neighbors.
This observation originates from the difference between induced subgraph isomorphism and embedding subgraph isomorphism.
VF2 and VF3 both adopt the definition of induced subgraph isomorphism, thus we must modify the original algorithms to satisfy our needs (Definition~\ref{def:graphisomorphism}).

\nop{
\Paragraph{Analysis}.
The implementation of VF3 only needs several arrays, with the array size proportional to graph size.
It does not maintain extra indices except for node classifiers.
Thus, the space complexity is very low: $O(|E(G)|)$.
As for performance,  \cite{VF3} claims that VF3 shows great superiority on dense graphs.
}


\subsection{CECI Algorithm} \label{sec:ceci}

CECI is similar to TurboISO\@: rewriting into NEC tree, finding candidate regions, etc.
However, it abandons the CR tree structure and proposes the CECI structure, which is similar to the CPI index.
Its main contribution is maintaining both tree edge candidates (TEC) and non-tree edge candidates (NTEC) for each NEC vertex $u^{\prime}$.
Due to the existence of NTEC, in Line~\ref{algcmd:join} we can use set intersection of TEC and NTEC.

\begin{figure}[htbp]   
	\centering
	\includegraphics[width=8cm]{\picfolder 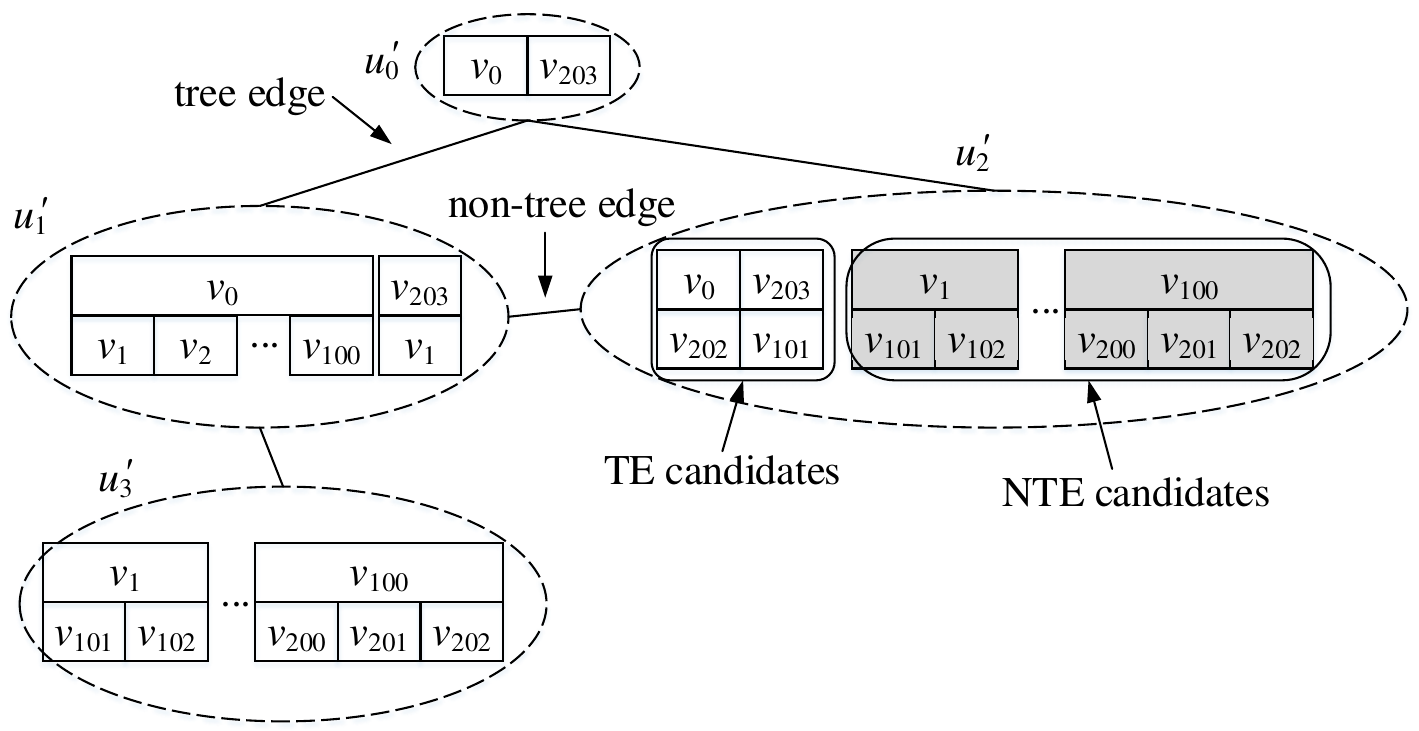}   
	\vspace{-0.1in}          
	\caption{CECI structure after BFS-based construction and Filtering}        
	\label{fig:ceci}   
\end{figure}

The CECI structure of Figure~\ref{fig:example} is shown in Figure~\ref{fig:ceci}, where $u_{i}^{\prime}$ is the NEC vertex in Figure~\ref{fig:turboiso}(a).
TEC is uncolored while NTEC is colored, for example, the colored area of Figure~\ref{fig:ceci} represents the set of NTE candidates of query edge $\overline{u_{1}^{\prime}u_{2}^{\prime}}$.
Within the circle of $u_{0}^{\prime}$, there are two candidate regions: $v_{0}$ and $v_{203}$.
In other query nodes, candidates are stored as pairs: the father and the corresponding children.
Taking $u_{2}^{\prime}$ as an example, a tree edge candidate is $(v_{0},v_{202})$, which means $v_{0}$ can be matched to the father $u_{0}^{\prime}$ and $v_{202}$ can be matched to the child $u_{2}^{\prime}$.
As for NTEC, the pair $(v_{1},\{v_{101},v_{102}\})$ means that $v_{1}$ can be matched to $u_{1}^{\prime}$, while $v_{101}$ and $v_{102}$ can be matched to $u_{2}^{\prime}$.

The CECI structure is built via a BFS traversal, then refined via a reversal-BFS traversal.
TEC and NTEC are both updated in the two traversal.
The matching order in verification is directly the BFS order.

\Paragraph{Analysis}.
After the CECI structure is built, there is no need to store adjacency list.
The space complexity is $O(|E(Q)|\times |E(G)|)$, which is higher than the CPI index of CFL-Match.
The time complexity of constructing CECI structure is $O(|E(Q)|\times |E(G)|)$.


\subsection{DAF Algorithm} \label{sec:daf}

Based on Algorithm~\ref{alg:generic}, DAF proposes several novel techniques for pruning.
For simplicity, we take the subgraph (of $Q$ in Figure~\ref{fig:example}) induced by $\{u_{0},u_{1},u_{2}\}$ as the example, which is a triangle.
Let $q$ be the new query graph.

\begin{figure}[htbp]   
	\centering
	\includegraphics[width=8cm]{\picfolder 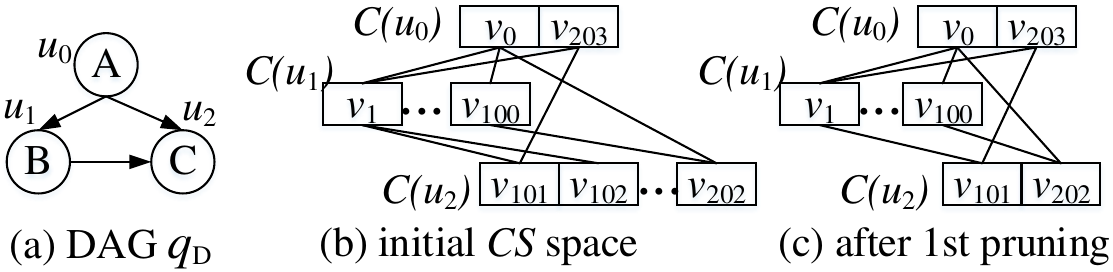}   
	\vspace{-0.1in}          
	\caption{Example of DAF algorithm}        
	\label{fig:daf}   
\end{figure}

First, instead of using a tree, a DAG $q_{D}$ is generated from $q$, which is directed and contains all the query edges.
An example is shown in Figure~\ref{fig:daf}(a).
By BFS from a selected root $r$, DAF directs all edges from upper levels to lower levels.
In the same level, directions are decided by vertex label frequency and vertex degree.
Vertices with more infrequent label and higher degree should come first.
The reverse DAG $q_{D}^{-1}$ can be obtained by reverse all edges in $q_{D}$.
During each iteration of filtering, a query DAG $q^{\prime}$ ($q^{\prime}$ can be $q_{D}$ or $q_{D}^{-1}$) is used.

Second, DAG-graph DP is performed for filtering, and candidates are stored in $CS$ space.
After filtering based on label and degree, the initial candidates $C(u)$ of each vertex $u$ are shown in Figure~\ref{fig:daf}(b).
To compute $C^{\prime}(u)$ by dynamic programming, DAF uses the following recurrence: \\
$v\in C^{\prime}(u) \iff v\in C(u)$ and $\exists v_{c}$ adjacent to $v$ such that $v_{c}\in C^{\prime}(u_{c})$ for every child $u_{c}$ of $u$ in $q^{\prime}$. \\
Figure~\ref{fig:daf}(c) shows the refined structure after first pruning by $q_{D}^{-1}$, which is similar to top-down filtering in CFL-Match.
Later, $q_{D}$ can be used for bottom-up refinement.
DAF alternates $q_{D}^{-1}$ and $q_{D}$  to optimize $CS$ until no changes occur.
Generally, 3 iterations are enough.

Third, based on DAG ordering, DAF adopts adaptive matching order.
Different paths in the search space may favor different matching orders.
The order is specially decided for each search path, thus it should be more accurate for pruning, though with higher computation cost for matching order selection.
In our example, the DAG ordering is unique, thus the matching order should be $u_{0}$->$u_{1}$->$u_{2}$.

\begin{figure}[htbp]   
	\centering
	\includegraphics[width=8cm]{\picfolder 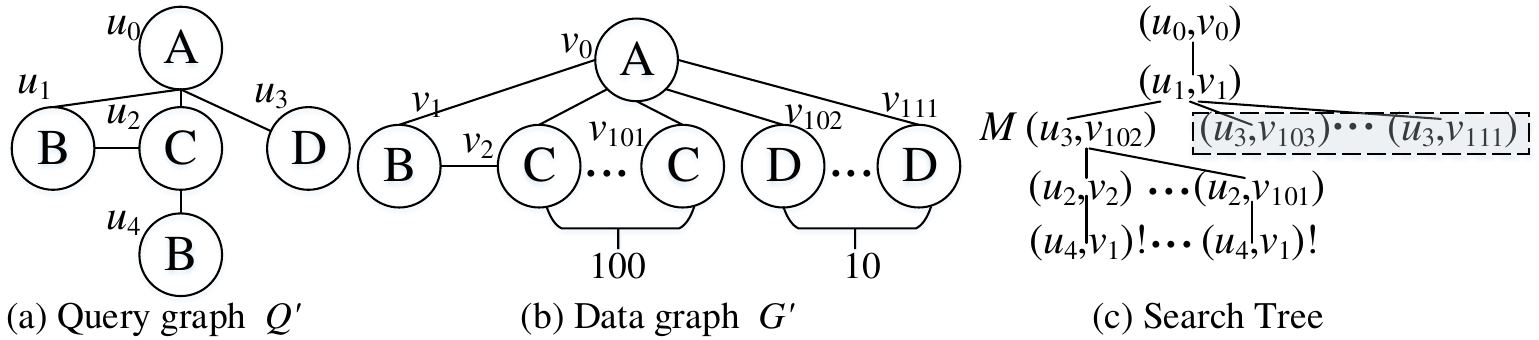}   
	\vspace{-0.1in}          
	\caption{Example of Failing Set}        
	\label{fig:newexample}   
\end{figure}

Fourth, DAF enhances pruning by failing sets.
In the search space, once we visit a new node $M$ (e.g., adding pair $(u,v)$), we need to explore the subtree rooted at $M$ and come back to $M$.
With the knowledge gained from the exploration of the subtree rooted at $M$, we may have chance to prune out the siblings (i.e., $u$ is mapped to other vertices) of node $M$.
For ease of presentation, we use a new example here, as shown in Figure~\ref{fig:newexample}.
After exploring subtree of $M=(u_{3},v_{102})$, the failing set $fs$ of node $M$ is acquired: $\{u_{0},u_{1},u_{2},u_{4}\}$.
During the exploration, all possible extensions are tried to map $u_{2}$ and $u_{4}$, but all attempts fail due to the conflict of $u_{1}$ and $u_{4}$.
However, $u_{3}$ is not involved in the conflict and not relevant to any failure.
This means that no matter how we change the mapping of $u_{3}$, it will never lead to a complete match.
Thus, the siblings of $M$ can be pruned, which are depicted with shadow box in Figure~\ref{fig:newexample}(c).

\Paragraph{Analysis}.
The time complexity of DAG-Graph DP is $O(|E(Q)|\times |E(G)|)$. 
DAF maintains candidates of all query edges in $CS$ space, thus the space complexity is also $O(|E(Q)|\times |E(G)|)$.


\section{Revisiting BoostISO} \label{sec:optimization}

BoostISO contains a set of preprocessing techniques on data graph that can be integrated into all existing subgraph isomorphism algorithms. 
However, some techniques may cause wrong answers or may be inefficient.
In this section, we revisit the techniques of BoostISO and analyze them in depth.

\subsection{Relationships and Techniques} \label{sec:boosttechniques}

The four kinds of relationships in data graph are as follows:

\begin{definition}\label{def:SC} \textbf{(Syntactic containment)}
	$\forall v_i, v_j \in G$, $v_i$ syntactically contains $v_j$ if $L(v_i)=L(v_j)$ and $N(v_i)\setminus \{v_j\} \subseteq N(v_j)\setminus \{v_i\}$, denoted $v_i \succeq v_j$.
\end{definition}

\begin{definition}\label{def:SE} \textbf{(Syntactic equivalence)}
	$\forall v_i, v_j \in G$, $v_i$ is syntactically equivalent to $v_j$ if they share the same neighbor set, denoted $v_i \simeq v_j$. The equivalent class of data vertices induced by SE is called syntactic equivalent class (SEC).
\end{definition}

\begin{definition}\label{def:QDN} \textbf{(Query-dependent neighbors)}
	The set of data vertices $\{v_i|v_i \in N(v), L(v_i) \in \{L_Q(u_i)|u_i \in N(u)\}\}$, denoted $QDN(Q,u,v)$.
\end{definition}

\begin{definition}\label{def:QDC} \textbf{(Query-dependent containment)}
	$\forall v_i$, $ v_j \in G$, if $L(v_i)=L(v_j)$ and $QDN(Q, u, v_j)-{v_i} \subseteq QDN(Q, u, v_i)-{v_j}$, we say $v_j$ QD-contains $v_i$ with respect to $u$ and $Q$, denoted $v_i \succeq_{(Q,u)} v_j$.
\end{definition}

\begin{definition}\label{def:QDE} \textbf{(Query-dependent equivalence)}
	if $v_i \succeq_{(Q,u)} v_j$ and $v_j \succeq_{(Q,u)} v_i$, we say $v_j$ is QD-equivalent to $v_i$ with respect to $u$ and $Q$, denoted $v_i \simeq_{(Q,u)} v_j$. 
\end{definition}

SC and QDC are partial order relations, while SE and QDE are equivalent relations of data vertices. 
SC and QDC can be represented in form of directed graph, with an arrow from $v_i$ to $v_j$ indicating $v_j \succeq v_i$.
We say $v_i$ is SC-Parent of $v_j$ and $v_j$ is SC-Child of $v_i$. 
For example, in data graph $G$ (Figure~\ref{fig:example}(b)), $L(v_1)=L(v_2)$ and $N(v_2)\subseteq N(v_1)$, so we have $v_1\succeq v_2$. 
Similarly, we have $v_{202}\succeq v_{201}$, $v_{202}\succeq v_{200}$, and $v_{200}\simeq v_{201}$.
Note that SC and SE relationships are irrelevant to specific queries, thus can be calculated offline.
On the contrary, QDC and QDE are calculated online. 
With these relationships, two techniques are proposed: data graph compression and dynamic candidate loading (DCL).

\Paragraph{Data Graph Compression}.
SE and QDE help reduce the size of data graph.
Data vertices with SE relationship can be viewed equivalently in terms of subgraph matching. 
If $v_i$ is syntactically equivalent to $v_j$ and $u$ can be matched to $v_i$, then we know that $u$ can also be matched to $v_j$. 
With this rule, we can merge $v_i$ and $v_j$ into a single node $h$ (called hypernode).
In this way, $V(G)$ is transformed into hypernode set $V_{sh}$.
Each hypernode $h$ of $V_{sh}$ corresponds to a SEC, which consists of several syntactically equivalent vertices of $G$.
By similar method, QDE can be exploited to merge hypernodes during query processing, thus further accelerates subgraph matching.
All QDE vertices of $h$ form the set $QDE-List(h)$.

\begin{figure}[htbp]   
	\centering
	\includegraphics[width=8cm]{\picfolder 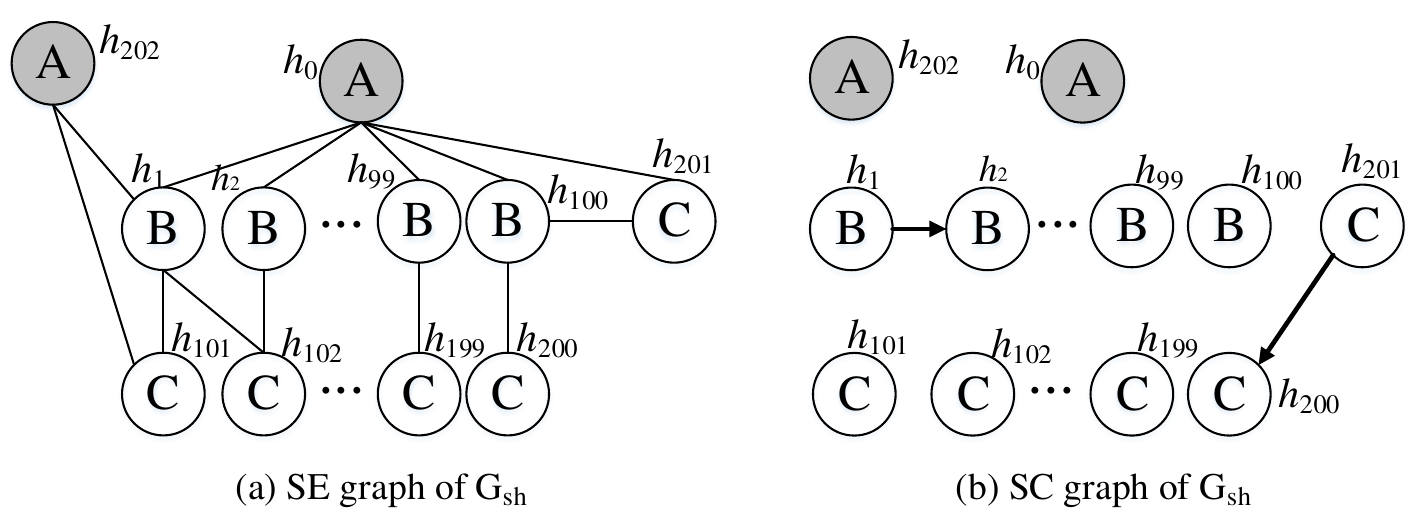}   
	\vspace{-0.1in}          
	\caption{Example of BoostISO}        
	\label{fig:boost}   
\end{figure}

\Paragraph{Dynamic Candidate Loading}.
SC and QDC help reduce the size of vertex candidate sets by setting matching order of data vertices.
If $h$ fails to match $u$, we know that the vertices which are syntactically or query-dependently contained by $h$ cannot be matched to $u$, either. 
For each $u$ in $V(Q)$, $C(u)$ is initialized with indegree-zero hypernodes in SC graph with respect to $Q$ and $u$. 
Once a hypernode is tested to successfully matched to $u$, namely true value is returned by the callee, such hypernode is removed from SC graph. 
Then new indegree-zero hypernodes may emerge and they need to be added to $C(u)$.
Finally, vertices in $C(u)$ have all been tested with no more vertices added. 
The matching order of data vertices in $C(u)$ is a topological sequence of SC graph. 
And the candidates that are syntactically contained by other hypernodes will always be tested later than its SC-Parents.
For example, in Figure~\ref{fig:boost} $h_{1}$ is tested before $h_{2}$ and $h_{201}$ is tested before $h_{200}$.

The original data graph $G$ is preprocessed to generate a hypergraph $G_{sh}$ before matching.
The hypergraph of Figure\ref{fig:example}(b) is shown in two parts: SE graph in Figure\ref{fig:boost}(a), and SC graph in Figure\ref{fig:boost}(b). 
In Figure~\ref{fig:boost}(a), $v_{200}$ and $v_{201}$ are syntactically equivalent and  merged into one hypernode $h_{200}$.
In contrast, $h_{201}$ only has one real vertex, i.e., $v_{202}$.
Let $R(h)$ be the real vertex set of $h$, we have $R(h_{200})=\{v_{200},v_{201}\}$ and $R(h_{201})=\{v_{202}\}$.
$v_{201}$ is syntactically contained by $v_{202}$, thus there is a directed edge from $h_{201}$ to $h_{200}$ in Figure~\ref{fig:boost}(b).

To conduct subgraph isomorphism search, BoostISO first finds all matches (called \emph{hyperembedding}s) of $Q$ in $G_{sh}$ following the semantics of subgraph homomorphism.
This matching process can be done by any existing subgraph isomorphism algorithm.
However, the original \emph{IsJoinable(...)} function (Line~\ref{algcmd:join0} of Algorithm~\ref{alg:generic0}) needs to be revised because two query vertices are allowed to be matched to the same hypernode.
The revised function is given in Algorithm~\ref{alg:join}.
For example, in Figure~\ref{fig:boost} the hyperembedding is $\{(u_{0},h_{0}),(u_{1},h_{100}),(u_{2},h_{201}),(u_{3},h_{200}),(u_{4},h_{200})\}$.
After acquiring all hyperembeddings, BoostISO transforms them to normal matches satisfying the semantics of subgraph isomorphism.
The transformation can be done by enumerating all combinations of real vertices, i.e., $R(h_{1})\times R(h_{2})...\times R(h_{n})$.
In our example, the enumeration occurs in $u_{3}$ and $u_{4}$: $\{v_{200},v_{201}\}\times \{v_{200},v_{201}\}$.
Duplicates are not allowed in this process, thus there are two normal matches left.

\nop{
\Paragraph{Analysis}.
To find all SC and SE relationships, each pair of data vertices needs to be tested, and all neighbors of the two vertices needs to be visited when testing a pair at worst case. 
Thus, $G_{sh}$ construction time is $O(|V(G)|^3)$. 
Similarly, the time cost of finding all QDC and QDE relationship for any given $u$ is also $O(|V(G)|^3)$. 
The number of hypernodes and edges in both SC and SE graph are no more than $|V(G)|$ and $|E(G)|$ respectively. 
}

\subsection{Error Analysis} \label{sec:boostdrawbacks}

In this section, we analyze the errors in BoostISO and give our solution.
After adding all techniques of BoostISO, there are two mistakes.
The first is that some valid matches are missed.
Though \cite{DAF}  also discovers this phenomenon, it does not explain the reason.
We find that DCL technique should be accused of this, which means SC and QDC relationships can not be leveraged.
Besides, using of QDE relationship can produce extra invalid matches, which is another mistake.

\Paragraph{Error of DCL technique}. 
In Figure~\ref{fig:example}, assuming $u_{0}$ and $u_{1}$ are matched to $h_{0}$ and $h_{100}$ respectively, $u_{2}$ can only be matched to $h_{201}$ while $u_{3}$ can be matched to $h_{200}$ and $h_{201}$.
If we match $u_{3}$ before $u_{2}$, the result will be empty.
$h_{200}$ is syntactically contained by $h_{201}$, thus $h_{201}$ is matched to $u_{3}$ first, which will fail when matching $u_{2}$ later.
Thus, $h_{200}$ will not be tried for $u_{3}$, though it can produce valid complete matches.
The inherent reason is that 1-hop neighborhood containment can not represent the global information.
Therefore, some valid matches are missed when using dynamic candidate loading.

\begin{algorithm}
    \small
	\caption{Revised IsJoinable function}
	\label{alg:join}
    \KwIn{query graph $Q$, hyper graph $G_{sh}$, partial match $f$, current considered pair ($u$,$h$)}
    \KwOut{\textit{true} if ($u$,$h$) can be added to $f$, \textit{false} otherwise}
\nl \ForEach{$u_{i}$ in $Q$}
{
\nl     \If{$u_{i}$ is not mapped by $f$}
    {
   \nl      \textbf{continue} \\
    }
  \nl   \If{$f(u_{i})\neq h$}
    {
\nl         \If{$\overline{u_{i}u} \in E(Q)$ and $\overline{f(u_{i})h}\notin E(G_{sh})$}
        {
\nl             \KwRet \textit{false} \\
        }
    }
\nl     \Else
    {
\nl         \If{$\overline{u_{i}u} \in E(Q)$ and $h.isClique=false$}
        {
\nl             \KwRet \textit{false} \\
        }
\nl         \If{$usedTimes(h)\geq |h|+\sum_{h_{i}\in QDE-List(h)}|h_{i}|$}
        {
\nl             \KwRet \textit{false} \\
        }
    }
}
\nl \KwRet \textit{true} \\
\end{algorithm}

\Paragraph{Error of QDE relationship}. 
Algorithm~\ref{alg:join} outlines the process of judging whether a new pair can be added to the partial match.
Assume that $v_{201}$ is eliminated from Figure~\ref{fig:example}.
In this case, there is no valid match and $R(h_{200})=\{v_{200}\}$.
Assume that the matching order  is $u_{0},u_{1},u_{2},u_{3},u_{4}$ and $u_{2}$ is already matched to $h_{201}$.
For $u_{3}$ and $u_{4}$ in Figure~\ref{fig:example}, $h_{201}$ can be viewed as a QDE vertex of $h_{200}$.
Thus, we have $|h_{200}|+\sum_{h_{i}\in QDE-List(h_{200})}{|h_{i}|}=|h_{200}|+|h_{201}|=2$.
Therefore, $u_{3}$ and $u_{4}$ can be both matched to $h_{200}$ according to Algorithm~\ref{alg:join}.
Thus, searching $Q$ on $G_{sh}$ can yield a hyperembedding $\{(u_{0},h_{0}),(u_{1},h_{100}),(u_{2},h_{201}),\\(u_{3},h_{200}),(u_{4},h_{200})\}$.
Later, we need to enumerate combinations.
For $u_{3}$ and $u_{4}$, all real vertices in $R(h_{200})\cup R(h_{201})$ need to be enumerated, because $h_{201}$ is QD-equivalent to $h_{200}$ on $u_{3}$ and $u_{4}$.
It seems there exists valid matches of searching $Q$ on $G$.
However, $h_{201}$ is already matched to $u_{2}$, and there is no unused real vertex left within $R(h_{201})$.
There is only one real vertex in $h_{200}$, thus the enumeration has no valid result.
Algorithm~\ref{alg:join} makes a mistake here, because it does not consider the consumption of $h$'s QDE vertices  otherwhere.
Generally, $h^{\prime}$ is QDE of $h$ on $u$, but it may not be QD-equivalent to $h$ on $u^{\prime}$.
If $h^{\prime}$ is also used for other query vertices, it can not be fully used as QDE for $u$.
In conclusion, some invalid matches may be produced when using QDE.

\Paragraph{Solutions of these errors}.
Due to the inherent restrictions of containment semantics, the error of DCL technique is impossible to correct.
Thus, SC and QDC relationships can not be used.
As for the error of QDE relationship, there are two possible solutions.
The first solution is to check and update the states of  all QDE vertices in each recursive call, which is not better than  withdrawing QDE.
The other solution is to verify each combination during final enumeration. 
But this is inefficient because many combinations are invalid and the results may be empty after verification.
If not using QDE, there must exist valid matches in enumeration.
Though a few duplicates may also exist, most combinations are valid. 
Besides, usage of DCL technique and QDE relationship can bring extra cost in both time and space because they need to maintain extra structures.
In conclusion, combining correctness and performance, among the four relationships, only SE should be  used for acceleration.

\section{Experiment} \label{sec:experiment}

We evaluate the state-of-the-art subgraph isomorphism algorithms QuickSI (abbreviated as QSI), GraphQL (abbreviated as GQL), TurboISO (abbreviated as TBI), CFL-Match (abbreviated as CFL), VF3, CECI and DAF, as well as their improved versions (QSI+, GQL+, etc) after adding optimizations of BoostISO.
Source codes are implemented in C++ and compiled with \textit{g++ 4.8.5} using \textit{-O2} flag.
All experiments are carried out on a workstation running CentOS 7 and equipped with Intel Xeon E5-2697 2.30GHz CPU and 188G host memory.
The matching process ends when finding the first $k=10^5$ matches or exceeding the time limit (10 minutes).
A query is claimed as \emph{solved} if it finishes within the time limit.
Note that there is no point if the  statistic exceeds the threshold.

GQL uses two techniques for filtering: neighborhood signature and pseudo isomorphism test, which is done by judging the containment of BFS trees.
We need to decide the hops ($k_{0}$) of neighborhood information used as signature.
Besides, the depth ($r$) of BFS tree needs to be tuned.
Different from \cite{SubgraphIsomorphismComparison2012}, in our context we set $k_{0}=1$ and $r=4$ for better performance.

As for QSI, in our implementation all filtering indices are not used.
Though the filtering indices proposed by \cite{QuickSI} are complex and useful, they only target at graph database.
For subgraph matching on a single data graph, these indices do not work.
Therefore, QSI does not have filtering phase and we only utilize the technique of  matching order selection.
As a result, the elapsed time of QSI is proportional to the number of recursive calls, which is verified by our experiments.

\Paragraph{Datasets}.
Our experiments use four real graphs: Yeast, Human, Email and DBLP, which are also used in \cite{SubgraphIsomorphismComparison2012, CFL-Match, DAF}.
The statistics are listed in Table \ref{tab:dataset}.  
The original Email and DBLP  do not have vertex label, thus we randomly assign 20 distinct labels to vertices.
The sparsity can be scored by average degree and lower average degree means high sparsity.
Thus, the ranking order of sparsity is DBLP>Yeast>Email$\gg$Human.


\begin{table}[htbp]
	\begin{threeparttable}
		\small
	\vspace{-0.1in}
		\centering
		\begin{tabular}{ccccccc}
			\toprule
            Name($G$) & $|V(G)|$ & $|E(G)|$ & $|\sum|$  & max-deg  & avg-deg \\
			\midrule
            Yeast & 3,112 & 12,519 & 179 & 168 & 8.0 \\
            Human & 4,674 & 86,282 & 88 & 771 & 36.9 \\
            Email & 36,692 & 183,831 & 20 & 1383 & 10.0 \\
            DBLP & 317,080 & 1,049,866 & 20 & 343 & 6.6 \\
			\bottomrule
		\end{tabular}
		\label{tab:dataset}
	\end{threeparttable}
	\caption{Statistics of Datasets}
	\vspace{-0.1in}
\end{table}

\Paragraph{Queries}.
Following the setting of \cite{CFL-Match}, we generate 8 query sets ($Q_{iS}$, $Q_{iN}$), where $i\in \{30,40,50,60\}$ for Yeast and $i\in \{10,20,30,40\}$ for others.
Each query set consists of 100 query graphs ($q_{0}$\textasciitilde$q_{99}$):
(1) $\forall q\in Q_{iS}$, it has $i$ vertices and is sparse (i.e., avg-deg($q$)<3);
(2) $\forall q\in Q_{iN}$, it has $i$ vertices and is non-sparse (i.e., avg-deg($q$)$\geq$3).
To generate a query graph $q$, we perform random walk over $G$ until $|V(q)|$ vertices are visited. 
All visited vertices and edges form a query graph. 

\begin{table}[htbp]
    \begin{threeparttable}
        \small
    \vspace{-0.1in}
        \centering
        \begin{tabular}{ccccc}
            \toprule
            Name($G$) & time(ms) & $|V(G_{sh})|$ & $|E(G_{sh})|$ & compress ratio \\
            \midrule
            Yeast & 77 & 3,040 & 12,435 & 10.0\%   \\
            Human & 516 & 2,939 & 34,389 & 59.0\%    \\
            Email & 14,057 & 26,320 & 166,947 & 12.4\%    \\
            DBLP &  519,546 & 275,451 & 875,006 & 15.8\%    \\
            \bottomrule
        \end{tabular}
        \label{tab:boost}
    \end{threeparttable}
    \vspace{-0.1in}
    \caption{Statistics of BoostISO}
\end{table}

\begin{figure*}[htbp]
    \centering
    \includegraphics[width=16cm]{\picfolder 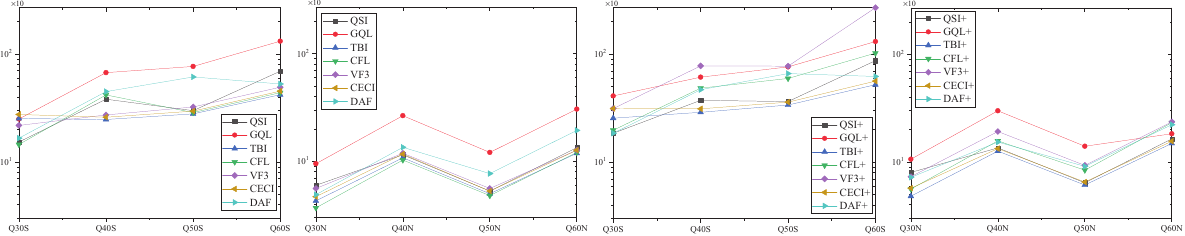}   
    \vspace{-0.1in}          
    \caption{Elapsed time (ms) on Yeast dataset}        
    \label{fig:yeastTime}   
\end{figure*}


\Paragraph{Measure}.
Regarding each query set, we report the average response time (ms) and peak memory consumption (KB).
Following \cite{DAF}, let $n$ be the minimum number of solved queries among all algorithms, only the $n$ least time-consuming queries are counted.
To ensure the meaning of average, we require that each algorthm $f$ must pass at least 10 queries on each query set $qs$; otherwise, $f$ is not \emph{workable} and  will not be compared on set $qs$.
The minimum value $n$ is computed on all workable algorithms.
Besides, the percentage of solved queries is also reported.

For BoostISO, its preprocessing time and size of $G_{sh}$ are given in Table \ref{tab:boost}.
Obviously, as the graph becomes larger, the preprocessing time of BoostISO increases by several orders of magnitude.
We use the formula $1-\frac{|V(G_{sh})|+|E(G_{sh})|}{|V(G)|+|E(G)|}$  to compute the compress ratio of BoostISO. 
The ranking order of compress ratio is Human$\gg$DBLP>Email>Yeast.
The memory cost of boosted versions is not presented because the relative comparison is similar to non-boosted versions.
Similarly, we do not depict the solved percentage of boosted versions because they do not show any advantage when compared to non-boosted solutions.

In later subsections, we will analyze the experimental results on each dataset one by one.
Analysis of each dataset occupies an entire subsection.
Elapsed time mainly consists of the filtering time and joining time.
More precise filtering is more costly, but yields smaller candidate sets.
The joining time is decided by the number of recursive calls and the cost of each call.
The number of recursive calls (also called backtracking steps), is mainly influenced by the sizes of candidate sets, the pruning techniques and the matching order.

\begin{figure}[htbp]   
    \centering
    \includegraphics[width=8cm]{\picfolder 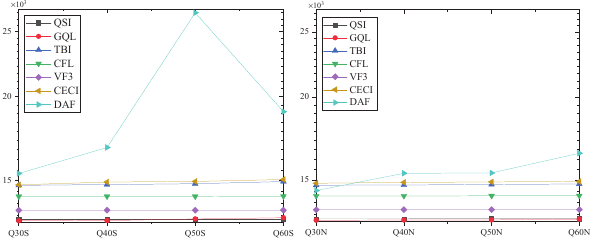}   
    \vspace{-0.1in}          
    \caption{Memory consumption (KB) on Yeast dataset}        
    \label{fig:yeastMem}   
\end{figure}

\subsection{Yeast Dataset} \label{sec:expYeast}

In this section, we analyze the experimental results on Yeast.
Since Yeast is much smaller than other datasets, we use larger queries for it.
When query size is bigger than 60, we can not generate enough queries from the data graph.
That is the reason why we prepare such query sets for Yeast.
In this experiment, all algorithms solve all queries successfully, i.e., the percentage of solved queries is always 100\%.

From Figure \ref{fig:yeastTime}, we can get some overall observations:

(1) As the query size grows larger, the time cost does not necessarily increase.
Though larger queries have more query vertices for backtracking, they generally have fewer matches, which may improve the probability of early pruning.
On dense query sets, the early pruning occurs more frequently.

(2) Comparing corresponding S/N sets (e.g., Q30S and Q30N), performance on non-sparse sets is much better.
As Yeast is a very sparse graph, denser queries mean stronger pruning power in both filtering and joining, thus they have smaller candidate sets and need fewer recursive calls.

(3) Boosted versions does not show much superiority on Yeast.
For some algorithms (like DAF and VF3), boosted versions are a little worser on all query sets.
Table \ref{tab:boost} shows that the compress ratio of BoostISO is the smallest on Yeast, which can help explain this phenomenon.
Boosted versions can bring little benefit, but add extra cost like enumerating equivalent data vertices and more costly \emph{IsJoinable} judgement.
That is why DAF+ performs worse than DAF though they have similar number of recursive calls.

Considering elapsed time of non-boosted solutions, GQL performs worst on all query sets.
Though the number of recursive calls is small for GQL, its performance is dragged down by heavy filtering phase.
Curves of other algorithms are very close to each other, and there is no single winner among them.
Overall, TBI and CFL can be viewed as the best two.
Though DAF proposes failing sets, its effect is limited here because the search space is not large enough.

\begin{figure}[htbp]
    \centering
    \includegraphics[width=8cm]{\picfolder 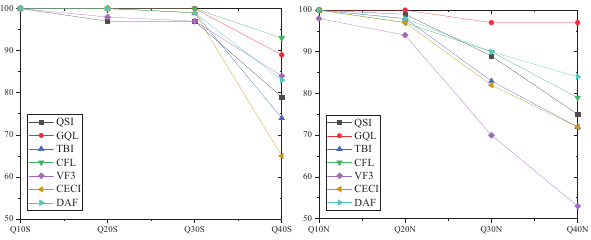}   
    \vspace{-0.1in}          
    \caption{Percentage of solved queries on Human dataset}        
    \label{fig:humanPSQ}   
\end{figure}

When it comes to boosted versions, GQL+ is not always the worst.
The gap between GQL+ and others is smaller than that on non-boosted solutions.
This reflects that BoostISO has greater impact on underperforming algorithms.
Differences of other algorithms are small.
The curve of VF3+ is higher than others.
On most queries, TBI+ and CECI+ performs the best among all boosted solutions.
Curves of other algorithms (QSI+, CFL+, DAF+) are crossed.

\begin{figure*}[htbp]
    \centering
    \includegraphics[width=16cm]{\picfolder 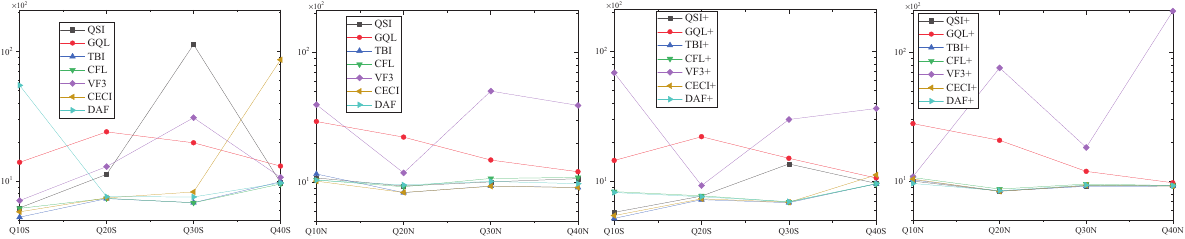}   
    \vspace{-0.1in}          
    \caption{Elapsed time (ms) on Human dataset}        
    \label{fig:humanTime}   
\end{figure*}


Figure \ref{fig:yeastMem} shows the comparison of memory consumption.
Among all non-boosted algorithms, obviously QSI and GQL claim the least memory cost.
The structures of QSI are simple and lightweight.
As for GQL, it needs to maintain neighborhood signatures, but in our setting ($k_{0}=1$) it is marginally not expensive.
DAF is the highest one because it maintains heavy data structures, i.e., $CS$ space and failing sets.
Especially, on Q50S, DAF shows a very high consumption.
However, on dense query sets, DAF shows much less memory cost because its strong filtering strategies greatly reduce the size of candidate sets.
Generally, the ranking order is QSI$\approx$GQL<VF3<CFL<TBI
$\approx$CECI<DAF.
The first three (QSI, GQL, VF3) are all tree-search algorithms, which does not keep auxiliary indices for candidates storage.
Thus, their memory consumption is the lowest.
As analyzed in Section \ref{sec:implementation}, CFL consumes less memory than CECI because it does not maintain candidates for non-tree edges.
TBI is a special case.
On the one hand, it merges NEC query vertices which helps lower the space cost of auxiliary indices.
On the other hand, its data structure (CR tree) is very costly.
Thus, its memory consumption is hard to be estimated.
On Yeast, TBI has similar memory consumption to CECI.

\begin{figure}[htbp]   
    \centering
    \includegraphics[width=8cm]{\picfolder 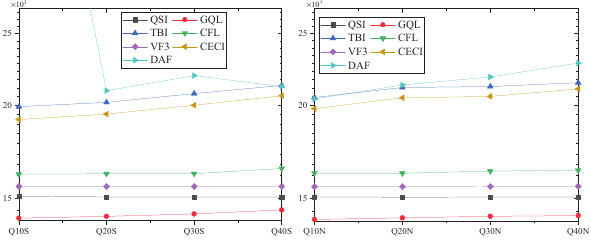}   
    \vspace{-0.1in}          
    \caption{Memory consumption (KB) on Human dataset}        
    \label{fig:humanMem}   
\end{figure}

\subsection{Human Dataset} \label{sec:expHuman}

In this section, we analyze the experimental results on Human.
The result of solved percentage is quite different from that on Yeast.
Figure \ref{fig:humanPSQ} gives the data.
The minimum value on each query set is (100, 96, 97, 65) for sparse sets, and (98, 87, 70, 50) for dense sets.
Generally, as query size grows, the solved percentage of all solutions drops.
Larger query size means longer processing time, thus more queries run timeout on larger query sets.
Comparing seven algorithms, the ranking order on dense sets is GQL>DAF>CFL>QSI
>TBI$\approx$CECI>VF3.
On sparse sets,  curves are crossed, but CFL is the best.
Starting from Q30S, the solved percentage drops more sharply.

\begin{figure}[htbp]   
    \centering
    \includegraphics[width=8cm]{\picfolder 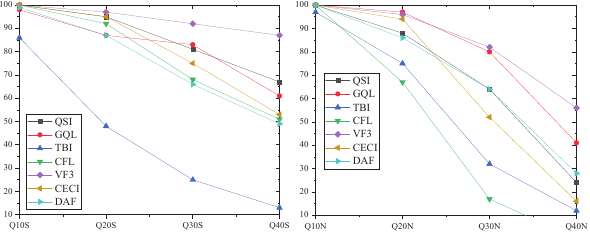}   
    \vspace{-0.1in}          
    \caption{Percentage of solved queries on Email dataset}        
    \label{fig:emailPSQ}   
\end{figure}

As the number of solved queries is quite different on various sets, we do not compare the performance between different query sets in this section.
Q20S has lower elapsed time than Q10S, but Q10S has higher solved percentage.
In this case, performance on Q20S is not better than Q10S because more time-consuming queries can be finished on Q10S.
For the same reason, Sparse sets should not be compared with dense sets.
For example, we needn't compare the performance on Q30S with that on Q30N.

Figure \ref{fig:humanTime}  shows the performance comparison on Human.
BoostISO has great impact on QSI, which yileds $5\times$ speedup on sparse query sets.
Besides, BoostISO also helps accelerate some extreme cases, such as QSI on Q30S, DAF on Q10S and CECI on Q40S.
The performance gain comes from the large compress ratio of BoostISO on Human, which helps reduce the number of recursive calls.
One exception is that VF3+ usually performs worse than VF3.
There exists some bad cases for VF3+, which terribly pulls high the averages.
For example, $q_{8}$ is the worst case for VF3+ in Q10S.

\begin{figure*}[htbp]   
    \centering
    \includegraphics[width=16cm]{\picfolder 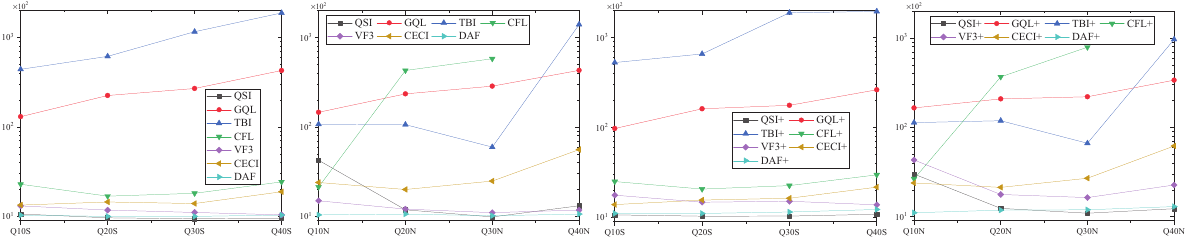}   
    \vspace{-0.1in}          
    \caption{Elapsed time (ms) on Email dataset}        
    \label{fig:emailTime}   
\end{figure*}


%

Comparing non-boosted solutions, TBI claims the best on both sparse and dense query sets.
CFL and CECI are the runner-up on sparse and dense query sets, respectively.
On sparse sets, other curves (except TBI and CFL) are messed up.
On dense sets, curves of VF3 and GQL are in higher position and the two curves are crossed.
Other curves are in lower position and they are very close to each other.
As for boosted versions, performance of VF3+ and GQL+ is the worst.
Differences of other boosted solutions are very small.

Figure \ref{fig:humanMem} shows the details of memory cost.
The ranking order is GQL<QSI<VF3<CFL<CECI<TBI<DAF.
Since Human is a dense graph, the sizes of candidate sets are usually large.
GQL has strong filtering power by neighborhood-based signature and pseudo-isomorphism test.
Thus, GQL has smaller candidate sets and lower memory consumption than QSI and VF3.
On Q10S, DAF consumes extremely large memory, 83MB.
There is a bad case for DAF in Q10S, i.e., the query $q_{7}$.
DAF performs terrible on $q_{7}$, thus all averaged measures are pulled too high by it.

\begin{figure}[htbp]   
    \centering
    \includegraphics[width=8cm]{\picfolder 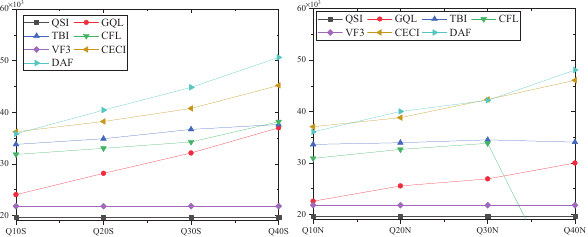}   
    \vspace{-0.1in}          
    \caption{Memory consumption (KB) on Email dataset}        
    \label{fig:emailMem}   
\end{figure}

\subsection{Email Dataset} \label{sec:expEmail}

In this section, we analyze the experimental results on Email.
The percentage of solved queries on Email is shown in  Figure \ref{fig:emailPSQ}.
The minimum value on each query set is (85, 47, 19, 13) for sparse sets, and (97, 67, 17, 11) for dense sets.
On sparse query sets, the ranking order is VF3>QSI>CECI>CFL>DAF>TBI.
GQL crosses several curves, and its percentage should be between QSI and DAF.
On dense sets, the ranking order is VF3>GQL>CECI,QSI,DAF>TBI
>CFL.
Curves of CECI, QSI and DAF are crossed, thus there is no absolute order of them.
On dense sets, CFL fails to answer all queries in Q40N.
The percentage drops when query size grows.
Generally, the dropping speed is higher on dense query sets than that on sparse query sets.

The performance comparison is shown in Figure \ref{fig:emailTime}.
On sparse query sets, the ranking order is QSI$\approx$DAF<VF3<CECI
<CFL<GQL<TBI.
The former five algorithms are very close to each other.
The superiority of QSI comes from its fantastic matching order selection, which greatly reduce the number of recursive calls.
On dense query sets, things are very different.
The ranking order of elapsed time can be represented as DAF<VF3<QSI<CECI
<TBI<GQL<CFL.
But this is not a strict order because there exist some crosses.
DAF is the best algorithm on both sparse and dense query sets.
As for boosted versions, they do not show any advantage on Email and the inner comparison is similar to that on non-boosted solutions.

As for the memory consumption on Email, Figure \ref{fig:emailMem} gives the details.
Generally, the ranking order is QSI<VF3<GQL<CFL<TBI
<CECI<DAF.
Note that CFL does not have value on Q40N.
On Email, the memory consumption of GQL is much higher than that of QSI and VF3.
As Email is much sparser than Human, the sizes of candidate sets are not quite different in QSI, VF3 and GQL, though GQL is equipped with strong filtering power.
Besides, since Email is much larger than Human, the memory cost of neighborhood-based signatures is high.
VF3 is the runner-up because it also needs to record extra information (e.g., the category information) for data graphs.
QSI is the most lightweight, because it does not have extra structures on data graphs, except for adjacency lists.

\begin{figure}[htbp]   
    \centering
    \includegraphics[width=8cm]{\picfolder 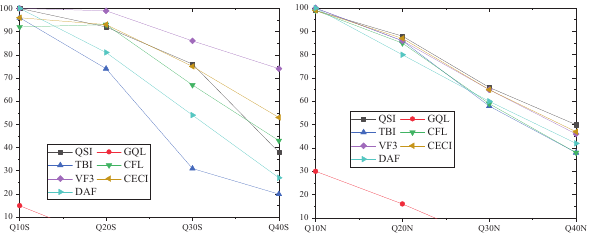}   
    \vspace{-0.1in}          
    \caption{Percentage of solved queries on DBLP dataset}        
    \label{fig:dblpPSQ}   
\end{figure}

\begin{figure*}[htbp]   
    \centering
    \includegraphics[width=16cm]{\picfolder 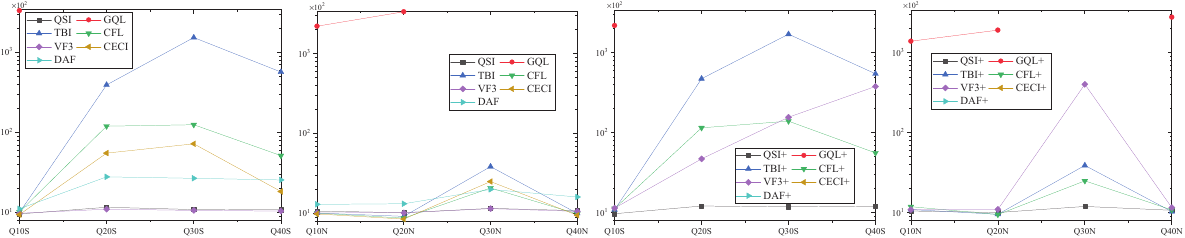}   
    \vspace{-0.1in}          
    \caption{Elapsed time (ms) on DBLP dataset}        
    \label{fig:dblpTime}   
\end{figure*}


\subsection{DBLP Dataset} \label{sec:expDBLP}

In this section, we analyze the experimental results on DBLP.
Figure \ref{fig:dblpPSQ} shows the percentage of solved queries on DBLP.
The minimum value on each query set is (15, 68, 31, 15) for sparse sets, and (30, 16, 39, 10) for dense sets.
On sparse query sets, the ranking order is VF3>CECI,CFL,QSI>DAF>TBI>GQL.
Among these, GQL is far below others.
On dense sets, QSI leads the way, but curves of all solutions except for GQL  are very close to each other.
Some algorithms fail to process some query sets.
GQL only passes Q10S, Q10N and Q20N, while GQL+ passes Q10S, Q10N, Q20N and Q40N.
In addition, CECI+ and DAF+ fail on all query sets.
In most cases, the percentage drops as the query size grows larger.
One exception is CFL, slightly rises from 92 on Q10S to 93 on Q20S.
On DBLP, the dropping speed does not change a lot.

Figure \ref{fig:dblpTime} shows the performance comparison on DBLP.
On sparse query sets, the ranking order of elapsed time is VF3$\approx$QSI
<DAF<CECI<CFL<TBI<GQL.
But on Q40S, CECI performs better than DAF.
On dense query sets, curves are messed up, but GQL is out of question the worst.
VF3 and QSI are the best two on most queries.
On Q10N, Q20N and Q40N, DAF is the last but one.
But on Q30N, TBI is the second worst.

Comparing boosted versions with non-boosted algorithms, only GQL+ is much better than GQL, while others do not have any superiority when compared to non-boosted solutions.
VF3+ even performs much worse than VF3, due to its large number of recursive calls.
This again illustrates BoostISO's significant influence on time-consuming cases.
But on more general cases, its effect is rather limited.

Targeting at boosted solutions only, the ranking order is QSI+<
VF3+,CFL+<TBI+<GQL+ on sparse query sets.
CFL+ is worse than VF3+ on Q20S, while on other sets CFL+ is better.
But on dense query sets, the ranking order is QSI+<CFL+<TBI+<VF3+
<GQL+.
The differences are only prominent on Q30N.

As for the memory consumption, the ranking order is QSI<VF3<
GQL<CFL<TBI<CECI<DAF.
However, there are some exceptions.
On Q10N, GQL is exactly between QSI and VF3.
Besides, DAF is better than TBI and CECI on Q10S, Q10N and Q20N.

\begin{figure}[htbp]   
    \centering
    \includegraphics[width=8cm]{\picfolder 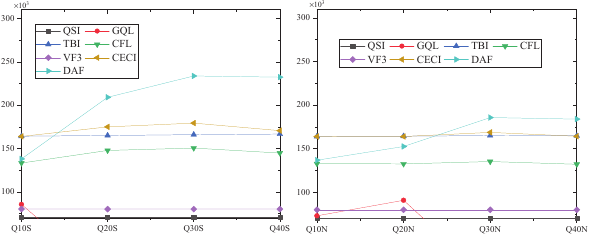}   
    \vspace{-0.1in}          
    \caption{Memory consumption (KB) on DBLP dataset}        
    \label{fig:dblpMem}   
\end{figure}



\section{Conclusion}\label{sec:conclusion}

In this paper, a fair comparison is provided for state-of-the-art subgraph isomorphism algorithms.
Besides, we clarify some mistakes in the usage of VF2/VF3 and BoostISO\@.
There is no single winner, but the overall top three solutions are VF3, QuickSI, DAF\@.
It is surprising that tree-search algorithms (VF3 and QuickSI) lead the way in both time performance and memory consumption.
Newer index-based algorithms all adopt heavy filtering and complex pruning techniques, but the final performance is not convincing enough.
Therefore, a better tradeoff between pruning ability and pruning overhead  should be explored in the future.
As for BoostISO, though the effect is rather limited, it can greatly improve many bad cases.
This indicates that we should decide whether to use BoostISO or not in a specific situation.
BoostISO performs well on very dense datasets, as it yields high compress ratio.
Besides, DAF shows superiority on dense queries and datasets because its failing set technique only works with large search space.
Our experiments and analyses will provide insights on the acceleration of subgraph isomorphism.
All source codes, along with experimental data, are released in \href{https://github.com/bookug/siep/}{Github}.

	\optionshow{}{
		\myproof{A Proof.}
	}
	
	
	
	
	\bibliographystyle{IEEEtran}
	\bibliography{siep}  
	
	
	
	
\end{document}